\documentclass[
    ,final            
  ]
  {aipproc}

\layoutstyle{6x9}
\usepackage{amsmath}

\begin{document}

\title{Azimuthally Sensitive Femtoscopy and $v_2$}


\author{Boris Tom\'a\v sik}{
  address={The Niels Bohr Institute, Blegdamsvej 17, 2100 Copenhagen \O, Denmark},
  address={and \'Ustav jadern\'e fyziky AV\v CR, 25068 \v Re\v z, Czech Republic},
  email={boris.tomasik@cern.ch}
}

\begin{abstract}
I investigate the correlation between spatial and flow anisotropy in 
determining the elliptic flow and azimuthal dependence of the HBT 
correlation radii in non-central nuclear collisions. 
It is shown that the correlation radii are in most 
cases dominantly sensitive to the anisotropy in space. In case of $v_2$, 
the correlation depends strongly on particle species. A procedure 
for disentangling the spatial and the flow anisotropy is proposed.
\end{abstract}

\maketitle


\section{Motivation}

In non-central nuclear collisions at RHIC energies, the resulting fireball
can exhibit anisotropy in both spatial shape and transverse expansion velocity
profile. They both influence the measured ``elliptic flow'' coefficient $v_2$
\cite{vol}. A question arises: how are they correlated in the determination
of $v_2$, i.e., which combinations of spatial and flow anisotropy lead to 
the same elliptic flow? 

On the other hand, dependence of HBT correlation radii on the azimuthal 
angle is also shaped by the two mentioned anisotropies. Therefore, the same 
question can be asked: how do the spatial anisotropy and 
transverse expansion flow anisotropy combine in  the 
$\phi$-dependence of correlation radii?

An analogical situation appears in determining the slopes of 
single-particle $p_T$-spectra. It is well known that they are determined by 
temperature and  transverse expansion velocity and that it is 
impossible to disentangle these two quantities from a single measured 
spectrum. There is, however, also the $M_T$-dependence of HBT
radii in which the correlation of temperature and transverse flow 
is qualitatively different from that  in the determination of spectra. 
Temperature and  transverse flow velocity then can be unambiguously 
measured from analysing both  spectra and  HBT radii.

A similar solution shall be sought here: can we disentangle spatial and
flow anisotropy in non-central collisions by analysing both $v_2$
and the azimuthally sensitive HBT radii?

Note that several statements have been made in  literature which are 
related to this programme. In \cite{st130} the STAR collaboration concluded
that it was impossible to determine spatial anisotropy just from 
the measurement of $v_2$ and a conjecture was made that  HBT analysis
would be able to gain such result. Two qualitatively different final 
states resulted from hydrodynamic simulations by Heinz and Kolb 
\cite{khplb} and the authors demonstrated the possibility to distinguish 
these states by HBT interferometry. Here I report on a systematic study 
of the interplay between spatial and flow anisotropy in framework of 
generalisations of the blast-wave model.


\section{An azimuthally anisotropic blast-wave model}

Instead of fully describing the used model I will 
just focus on those features which are important for this work and refer
the reader to literature for more detailed discussion \cite{retiere,asym}.
Suffice it to say that the fireball is thermalised with a temperature $T$ 
and exhibits longitudinally boost-invariant expansion. Its transverse 
profile is ellipsoidal and the emission function is 
\begin{equation}
S(x,p) \propto \Theta(1-\tilde r) \, , \qquad 
\tilde r = \sqrt{\frac{x^2}{R_x^2} + \frac{y^2}{R_y^2}}\, ,
\end{equation}
where $R_x$ and $R_y$ are the two  transverse radii, in and out-of-plane,
respectively. They can be parametrised with the help of a spatial 
anisotropy parameter $a$
\begin{equation}
R_x = a\, R\, , \qquad  R_y = \frac{R}{a}\, .
\end{equation}
Thus an out-of-plane elongated source is characterised by $a<1$, whereas
for an in-plane elongated source we have $a>1$. 

\begin{figure}
  \includegraphics[height=.25\textheight]{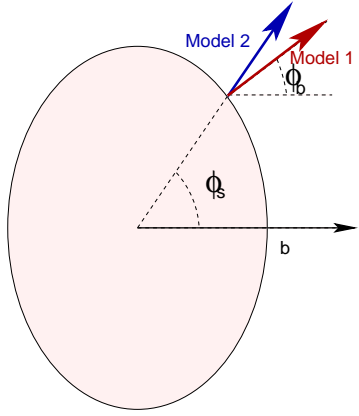}
  \caption{The two different models for transverse expansion 
    velocity used here.
  \label{f:models}}
\end{figure}

The transverse expansion velocity also depends on the azimuthal angle. The
velocity is given as 
\begin{equation}
v_\perp = \tanh \rho(\tilde r,\, \phi)\, .
\end{equation}
We shall have a closer look at two models which differ in the azimuthal 
variation of the velocity. In {\em Model 1} \cite{retiere} the velocity 
is always perpendicular to a surface given by $\tilde r = \mbox{const}$. 
This direction together with the reaction plane defines the azimuthal angle
$\phi_b$, as illustrated in Figure~\ref{f:models}. 
The transverse rapidity
\begin{equation}
\rho(\tilde r,\phi) = \tilde r \, \rho_0 \, (1 + \rho_2\, \cos(2\phi_b))\, ,
\end{equation}
where the parameter $\rho_0$ measures the radial flow and $\rho_2$ 
is the flow anisotropy parameter. As the velocity is perpendicular 
to the surface of the fireball, this model resembles the expansion profile 
early in the fireball evolution: the direction of velocity coincides 
with acceleration which in turn is given by the pressure gradient.

In {\em Model 2} the transverse expansion velocity is directed radially 
and varies with the usual azimuthal angle, which is denoted as $\phi_s$ here
\begin{equation}
\rho(\tilde r , \phi) = \tilde r \, \rho_0 \,(1+\rho_2 \,\cos(2\phi_s))\, .
\end{equation}


\section{The elliptic flow}

Recall that $v_2$ is defined as the second Fourier 
coefficient of the azimuthal dependence of  spectrum
\begin{equation}
P_1(p_T,\phi) = \left . \frac{d^3N}{p_T\, dp_T\, dy\, d\phi} \right |_{y=0} = 
\frac{1}{2\pi} \, \left . \frac{d^2N}{p_T\, dp_T\, dy}\right |_{y=0} 
\left (1 + 2 v_2(p_T)\cos(2\phi) + \dots \right )\, .
\end{equation}
It can be calculated in the two used models and the result reads
\cite{asym}
\begin{equation}
v_2 = \frac{\int_0^1 d\tilde r \, \tilde r \int_0^{2\pi} d \phi\, 
\cos(2\phi)\, J(\phi)\, K_1(a)\, I_2(b)}{%
\int_0^1 d\tilde r \, \tilde r \int_0^{2\pi} d \phi\, 
J(\phi)\, K_1(a)\, I_0(b)} \, ,
\end{equation}
where the arguments of the Bessel functions are 
$a = m_T\, \cosh\rho(\tilde r,\phi)/T$ and 
$b = p_T\, \sinh\rho(\tilde r, \phi)/T$. The {\em only} difference
between the two models appears in the Jacobian $J(\phi)$
\begin{subequations}
\begin{eqnarray}
\mbox{Model 1:}\qquad J(\phi) & = & (a^2\cos^2\phi + a^{-2}\sin^2\phi)\, ,\\
\mbox{Model 2:}\qquad J(\phi) & = & (a^{-2}\cos^2\phi + a^{2}\sin^2\phi)\, .
\end{eqnarray}
\end{subequations}
From these relations it is obvious that the two models lead to the same 
$v_2$ if they are related by transformation $a\to a^{-1}$. In other words, 
one in-plane and another out-of-plane source give the same $v_2$. This 
is an {\em 
analytic illustration of the claim that it is impossible to determine 
even the qualitative type of spatial anisotropy just from measurement of 
$v_2$}.

\begin{figure}
  \includegraphics[height=.39\textheight]{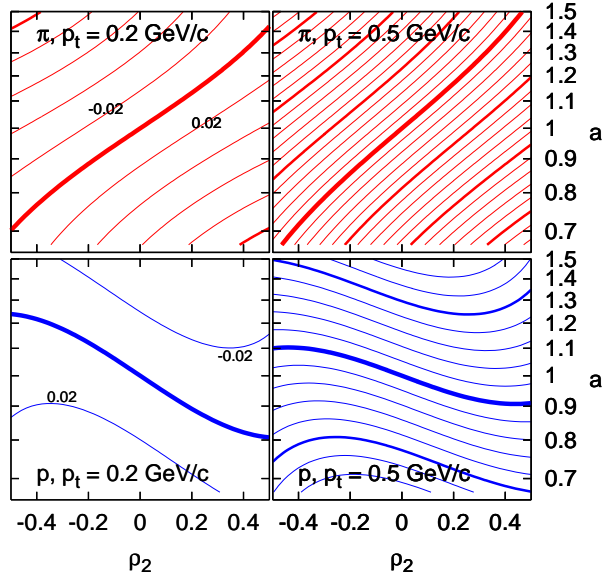}
  \caption{Elliptic flow $v_2$ calculated with Model 1 for pions (upper row) 
    and protons (lower row) with $p_T = 0.2\,\mbox{GeV}/c$ (left column)
    and 0.5~GeV/$c$ (right column). The used values $T=100\, \mbox{MeV}$ and
    $\rho_0 = 0.88$. Thickest contour lines show where $v_2$ vanishes; 
    consecutive lines correspond to steps by 0.02.
  \label{f:vdva}}
\end{figure}
Now we can look on the correlation between flow and spatial anisotropy and 
study it only for Model 1, since results for the other Model are obtained
simply by substitution $a \to a^{-1}$. In Figure~\ref{f:vdva} we see that 
the correlation between $a$ and $\rho_2$ strongly depends on the particle 
species. Hence, here is a strategy for determining both $a$ and $\rho_2$: 
first determine the temperature and radial flow coefficient $\rho_0$ from 
azimuthally integrated spectra. Their dependence on azimuthal 
anisotropies was shown to be small \cite{retiere}.  Then measure $v_2$ for 
at least two particle species and obtain $a$ and $\rho_2$. Of course, 
this procedure assumes that we know which model to use for the analysis. 
This leaves an open question which is to be answered by correlation 
measurement.


\section{Azimuthally sensitive HBT}

In non-central collisions, the HBT correlation radii can be measured as 
a function of the azimuthal angle $\phi$. We shall focus mainly on the two 
transverse radii $R_o$ and $R_s$ and 
decompose their azimuthal angle dependence as \cite{Heinz:2002au,twrev}
\begin{subequations}
\label{FDhbt}
\begin{eqnarray}
R_o^2(\phi) & = & R_{o,0}^2 + 2 R_{o,2}^2 \cos 2\phi + \dots\\
R_s^2(\phi) & = & R_{s,0}^2 + 2 R_{s,2}^2 \cos 2\phi + \dots \, .
\end{eqnarray}
\end{subequations}
The individual terms of these decompositions are obtained as various 
combinations of space-time variances taken with the emission function
\cite{Heinz:2002au,twrev}. Because we are rather interested in the 
oscillation of the radii and not so much in their absolute size, we
shall look at the normalised oscillation amplitudes $R_{i,2}^2/R_{i,0}^2$
\cite{retiere}%
\footnote{In fact, Reti\`ere and Lisa realised in \cite{retiere}
that because it also includes time contributions, $R_{o,0}^2$ is not a
good normalisation quantity and so used $R_{s,0}^2$ to normalise {\em all}
of $R_{o,2}^2,\, R_{s,2}^2$, and $R_{ol,2}^2$. This is not done here.}.
They are sensitive to $a$ and $\rho_2$, but less sensitive to $R$ and  
$\rho_0$.

\begin{figure}
  \centerline{{%
      \hspace*{-1.2cm}
    \begin{minipage}{7.9cm}
      \includegraphics[width=7.9cm]{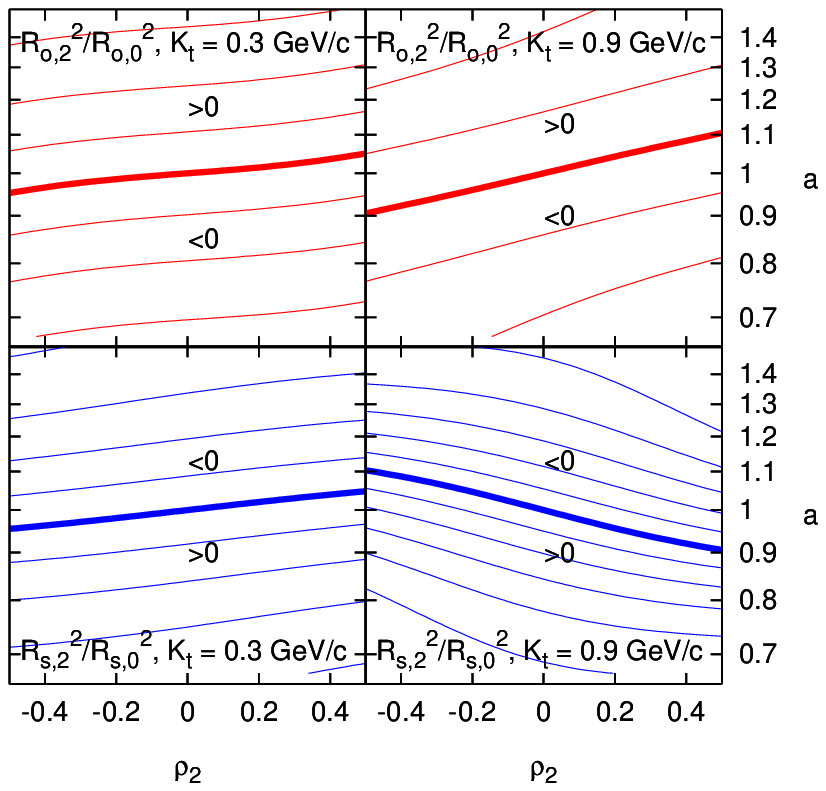}%
      \vspace*{-0.6cm}
      \centerline{Model 1}
      \centerline{}
    \end{minipage}
    \begin{minipage}{7.9cm}
      \includegraphics[width=7.9cm]{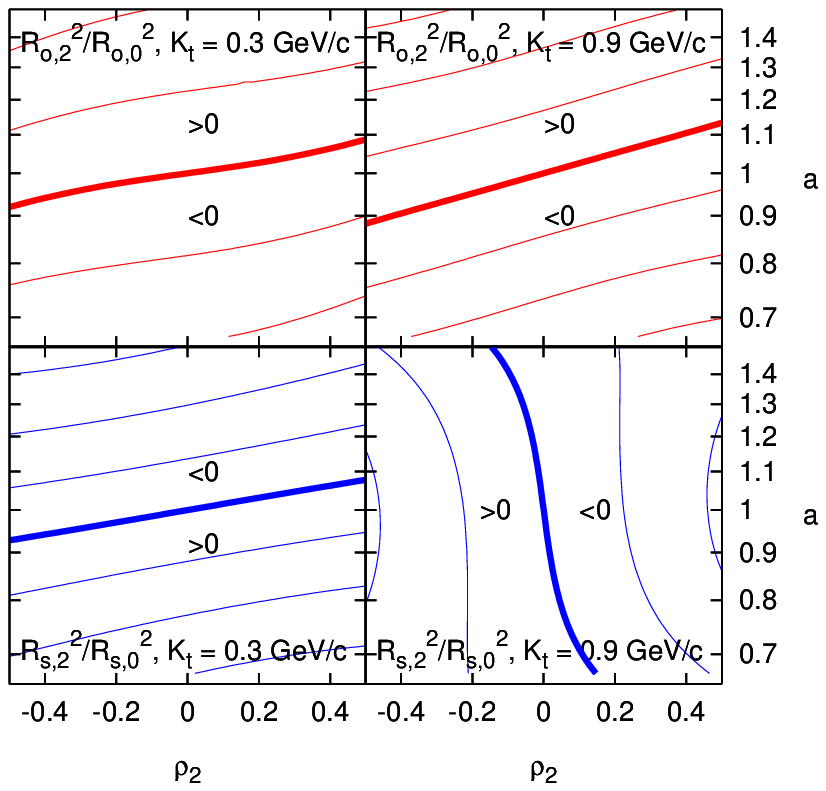}%
      \vspace*{-0.6cm}
      \centerline{Model 2}
      \centerline{}
    \end{minipage}}}
  \caption{Normalised oscillation amplitudes $R_{o,2}^2/R_{o,0}^2$ (upper row)
    and $R_{s,2}^2/R_{s,0}^2$ (lower rows) as functions of $a$ and $\rho_2$ 
    calculated for $T = 0.1\, \mbox{GeV},\, \rho_0 = 0.88,\, 
    R = 9.41\, \mbox{fm},\, \tau_0 = 9\, \mbox{fm}/c,\, 
    \Delta\tau = 1\, \mbox{fm}/c$ (see \cite{asym} for definitions of all
    parameters). Left columns show results for $K_T = 0.3\, \mbox{GeV}/c$,
    right columns correspond to $K_T = 0.9\, \mbox{GeV}/c$. Thickest lines
    show where the second-order oscillation terms vanish, other contours 
    are set in steps of 0.1. 
  \label{f:hbtvar}}
\end{figure}
From Figure~\ref{f:hbtvar} we conclude that the azimuthal oscillations 
of the HBT correlation radii are mainly shaped by the {\em spatial} anisotropy 
parameter $a$. Dependence on flow anisotropy is weaker, with the 
only exception of $R_s^2$ at high $K_t$ in Model 2 which is determined mainly 
by flow. This confirms the statement that the azimuthal dependence of 
correlation radii follows mainly the spatial anisotropy, especially at 
low $K_t$. This has been shown here in framework of two models. It would 
be natural to expect this behaviour to be valid in general. It can be spoilt 
by very strong flow gradients which differ by much in in-plane
and out-of-plane directions. A questions arises, however, whether 
large enough difference of the flow gradients is realistic.

\begin{figure}
  \includegraphics[height=.35\textheight]{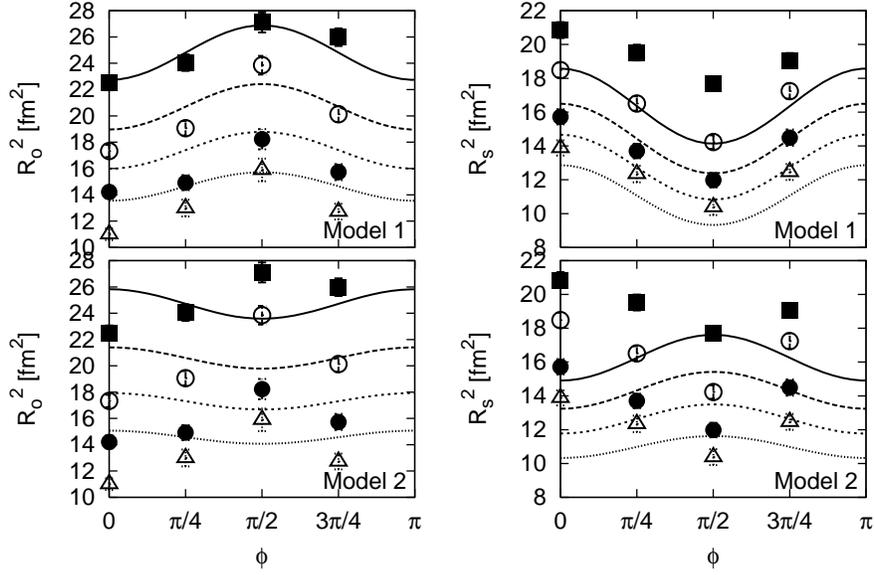}
  \caption{Comparison of the dependence of $R_{o}^2$ and 
    $R_{s}^2$ on azimuthal angle $\phi$ with the data
    measured by STAR collaboration (Au+Au at 200~$A$GeV, centrality class
    20--30\%) \cite{STARdata}. The curves
    and data points correspond from top to bottom to 
    $K_T = 0.2,\, 0.3,\, 0.4,\, 0.52\, \mbox{GeV}/c$. Parameters of the 
    models are: $T = 0.12\, \mbox{GeV}, \, \rho_0 = 0.99,\, \rho_2 = 0.035,\,
    R = 9.41\, \mbox{fm},\, \tau_0 = 5.02\, \mbox{fm}/c,\, 
    \Delta\tau=2.9\, \mbox{fm}/c$, and $a = 0.946$ (Model 1) or
    $a = 1.057$ (Model 2). Both models were tuned to fit $v_2(p_T)$ 
    \cite{Tomasik:2004bp}.
  \label{f:hbtfit}}
\end{figure}

The two investigated models exhibit similar dependence on the spatial 
anisotropy parameter $a$ when focusing on the oscillation of HBT radii. 
Recall, however, that they were related by transformation $a\to a^{-1}$ 
when reproducing the same $v_2$. Therefore, two different models which 
both reproduce $v_2$ measurement will behave differently when fitting 
the azimuthal dependence of HBT radii. This is illustrated in 
Figure~\ref{f:hbtfit}. 
%
%
Both models used in this figure fit measured the $v_2$ 
for pions and protons well. However, while Model 1 reproduces the 
RHIC data qualitatively well, Model 2 leads to the phase of oscillation 
just opposite to data \cite{STARdata}. 

Thus we conclude that among the two models used in this study, Model 1
seems to correspond to RHIC data, whereas Model 2 is clearly ruled out.
This does not disqualify it, however, from future applications at 
the LHC where possibly longer lived fireballs could be produced which 
will develop a different transverse flow pattern. 


\section{Conclusions}

It has been demonstrated analytically that 
one cannot disentangle spatial and flow anisotropy of the fireball just 
from a measurement of $v_2$. I also demonstrated that, at least for two 
classes of models,  the azimuthal dependence of correlation radii 
reflects the type of spatial anisotropy the source actually exhibits. 

Thus I can propose the following (schematic) procedure for disentangling 
$a$ and $\rho_2$: first measure the azimuthal dependence of HBT radii and 
determine the spatial anisotropy $a$. Then, with that $a$ try to reproduce
$v_2$ for more species. Since for different species $a$ and $\rho_2$ 
are correlated in different ways, this should lead to unique pair 
of the anisotropy parameters.


\section{Acknowledgements}

This research and presentation were supported by a Marie Curie Intra-European 
Fellowship within the 6th European Community Framework Programme.


\end{document}